\documentclass{aa}  
\usepackage{graphicx}
\usepackage{chemarr}
\usepackage{txfonts}
\usepackage{amsmath}
\usepackage{natbib}

\begin{document}

     \title{ Statistical universal branching ratios for cosmic ray dissociation, photodissociation, and dissociative recombination of the C$_{\rm n=2-10}$, C$_{\rm n=2-4}$H and C$_3$H$_2$ neutral and cationic species}

     \subtitle{ }

     \author{M. Chabot\inst{1}, T. Tuna\inst{1}, K. B\'eroff \inst{2}, T. Pino\inst{2}, A. Le Padellec\inst{4}, P. D\'esequelles\inst{3}, G. Martinet\inst{1}, V. O. Nguyen-Thi\inst{5}, Y. Carpentier\inst{2}, F. Le Petit\inst{6}, E. Roueff\inst{6}, \and V. Wakelam\inst{7,8}}
     \offprints{\email{chabot@ipno.in2p3.fr}}
     \institute{
      Intitut de Physique Nucl\'eaire d'Orsay,IN2P3-CNRS and Universit\'e Paris-Sud, 91406 Orsay cedex, France \and
      Institut des Sciences Mol\'eculaires d'Orsay, CNRS and Universit\'e Paris-Sud, 91405 Orsay cedex, France \and
      Centre de Spectroscopie Nucl\'eaire et de Spectrometrie de Masse, Universit\'e Paris-Sud and IN2P3-CNRS, 91405 Orsay cedex, France \and
      Centre d' \'Etude Spatiale des Rayonnements, Universit\'e  Paul Sabatier, UMR5187, 31028 Toulouse Cedex 9, France \and
      Laboratoire de Chimie Physique, Universit\'e Paris-Sud and CNRS, 91405 Orsay Cedex, France \and
      Laboratoire Univers et Th\'eories, CNRS et Observatoire de Paris, Place J. Janssen, 92195 Meudon cedex, France\and
     Universit\'e de Bordeaux, Observatoire Aquitain des Sciences de l'Univers, BP 89, F-33271 Floirac Cedex, France \and
     CNRS, UMR 5804, Laboratoire dAstrophysique de Bordeaux, BP 89, F-33271 Floirac Cedex, France 
    }

     \date{Received 18 May 2010 / Accepted 3 August 2010}
     
     \abstract
     { Fragmentation$-$branching ratios of electronically excited molecular species 
  are of first importance for the modeling of gas phase interstellar chemistry. Despite  experimental and theoretical efforts that have been done during the last two decades there is still a strong lack of detailed information on those quantities for many molecules such as C$_{n}$, C$_{n}$H or C$_3$H$_2$.}
     {Our aim is to provide astrochemical databases with more realistic branching ratios for C$_{n}$ ($n$=2 to 10), C$_{n}$H ($n$=2 to 4), and C$_3$H$_2$ molecules that are electronically excited either by dissociative recombination, photodissociation, or cosmic ray processes, when no detailed calculations or measurements exist in literature.}
     {High velocity collision in an inverse kinematics scheme was used to measure the complete fragmentation pattern of electronically excited C$_{n}$ ($n$=2 to 10), C$_{n}$H ($n$=2 to 4), and C$_3$H$_2$ molecules. Branching ratios of dissociation where deduced from those experiments. The full set of branching ratios  was used as a new input in chemical models and branching ratio modification effects observed in astrochemical networks that describe the dense cold Taurus Molecular Cloud-1 and the photon dominated Horse Head region.}
     {The comparison between the branching ratios obtained in this work and other types of experiments showed a good agreement. It was interpreted as the signature of a statistical behavior of the fragmentation. The branching ratios we obtained lead to an increase of the C$_3$ production together with a larger dispersion of the daughter fragments. The introduction of these new values in the photon dominated region model of the Horse Head nebula increases the abundance of C$_3$ and C$_3$H, but reduces the abundances of the larger C$_{n}$ and hydrocarbons at a visual extinction A$_{\rm V}$ smaller than 4.}   
    {We recommend astrochemists to use these new branching ratios. The data published here have been added to the online database KIDA (KInetic Database for Astrochemistry, http://kida.obs.u-bordeaux1.fr).}
    

     \titlerunning{Statistical Universal Branching Ratios}
     \authorrunning{Chabot et al.}

     \maketitle

\section{Introduction}

Carbon clusters and highly unsaturated hydrocarbons are among the  molecules most often observed in the Inter Stellar Medium (ISM) (http://astrochemistry.net).  They are present in diffuse clouds \citep{1982ApJ...254..108H, 2001ApJ...553..267M, 2002A&A...395..969G, 2002A&A...384..629R}, in dense clouds \citep{1999ApJ...518..740B, 2001ApJ...552..168F, 2000ApJ...542..870D}, in circumstellar envelopes \citep{2002ApJ...580L.157C}, protostar envelopes \citep{2008ApJ...672..371S}, planetary nebula \citep{2001ApJ...546L.123C}  and in photon dominated regions (PDR) \citep{2004A&A...417..135T}. They are also present around evolved (carbon) stars \citep{1988Sci...241.1319H},  in the Titan ionosphere  \citep{2008P&SS...56...67L} or in comet commae \citep{2005A&A...442.1107H}. Since their discovery in the early 70’s, many studies have been devoted to those species \citep{VanOrden1998}, mostly on their structural and spectroscopic properties. The spectroscopic studies were pushed mainly by the dramatically increasing performances of the observation capabilities, while for the structural studies it was also driven by the increasing use of carbon$-$based materials. 
      
      The origin of carbon clusters and highly unsaturated hydrocarbons and their abundances in the ISM objects is still a puzzling question. It is thought that they may come from gas$-$phase synthesis as well as be released into the gas$-$phase of carbonaceous solid materials that are present in the ISM. Therefore the detailed investigation of their origin and abundances requires homogeneous and heterogeneous chemical reactions to be explored. Whatever the carbon reservoir, those molecules are processed in the gas phase by neutral$-$neutral and ion$-$neutral collisions as well as cosmic rays (CR), ultra violet (UV) photons and dissociative recombination (DR) from collisions with the thermal electrons. These last three  mechanisms lead to highly electronically excited species that may undergo fragmentation. The detailed dynamics and the fragmentation channels have thus to be investigated to enable their incorporation  into various astrochemical models.
      
      In gas phase chemistry, chemical networks for carbon clusters and highly unsaturated hydrocarbons synthesis have been proposed for a large variety of astrophysical conditions \citep{2000MNRAS.316..195M, 1990ApJ...351..222J, 1977ApJS...34..405B, 1989ApJS...69..271H, 2000ApJS..126..427T, 2007A&A...467..187R, 1976Ap&SS..39....9G}. Nevertheless, despite the strong theoretical and experimental efforts in structure and spectroscopy, many of the reaction rates still remain uncertain and may consequently limit the confidence in chemical models. 

It is convenient to write the reaction rate coefficient for a reaction A+B $ \rightarrow$ C+D as the product of a total reaction rate coefficient $k_{tot}$  and a branching ratio factor (BR)

\begin{equation}
\rm k[A+B \rightarrow C+D] =  k_{tot}[A+B \rightarrow (AB)^\ast] \times  BR[(AB)^\ast \rightarrow C+D] .
\end{equation}

      The reliability of the reaction rate may  be improved when the total rate and BR are independently predicted.  Most of the reaction rates are sensitive to the surrounding medium (temperature and photon spectrum for instance). In Eq. (1) this dependency is implicitly included in the total rate,  whereas BRs are assumed to be constant. This assumption of constant BRs may however be inappropriate for some reactions as for example, neutral$-$neutral or ion$-$neutral reactive collisions where energy barriers can be present. For electronic excitation mechanisms followed by dissociation that occur in the ISM, the intermediate complex AB is appropriately described by an ensemble of molecules prepared in various highly exited electronic states. It has been pointed out several times that this physical situation fits the requirements of statistical theory concepts very well \citep{1978ApJ...222..508H, 1989A&A...213..351L, 1995IJMSI.149..321B}.  In the particular case of small species, non$-$statistical mechanisms need also to be considered because the density of states is generally not high enough, even if high$-$energy excitations are involved. Non$-$statistical behavior may also arise in rare photodissociation situations, where a dissociative state is resonant with a discrete line from a local photon source \citep{1988rcia.conf...49V}.

      In the first part of this paper we will report on measured BRs after electronic excitations that take place in a high$-$velocity collision experiment. In the second part we will introduce through the comparison between the different processes that are responsible for the electronic excitations in the ISM,  the idea of statistical universal BR for cosmic$-$ray$-$induced dissociation, photodissociation, and dissociative recombination processes. We will then propose to correct BRs of current online databases (OSU \footnote{http://www.physics.ohio-state.edu/~eric/research.html}, UMIST \footnote{http://www.udfa.net/}) when they are not resulting from measurements or detailed calculations. In a last part we will observe the influence of the new branching ratios on two typical astrochemical models that are aimed to simulate the dense clouds and the Horse Head PDR. 

\section{ Fragmentation of C$_{\rm n=2-10}$, C$_{\rm n=2-4}$H and C$_3$H$_2$ molecules that are electronically excited by high$-$velocity collision}
  
\subsection{Experimental set-up}
The experimental set-up has already been described in detail in previous papers \citep{2002NIMPB.197..155C,2008JChPh.128l4312T} and only a brief description is given here. Tandem MP (15 MVolts) accelerator at the Institut de Physique Nucléaire d'Orsay produced molecular cationic beams of several MeV of energy \citep{1996NIMPA.382..348W}. After magnetic analysis and collimation, singly charged molecules collided at high velocity (few a.u.) with a single target atom. Half a meter downstream, a parallel$-$plates electrostatic deflector analyzed  parents and fragments with respect to their q/m ratio.  A set of silicon detectors intercepted  all the trajectories in a dedicated chamber.  It is because of their high velocity that parent and fragments may be detected with silicon detectors. This type of detector enables the measurement of the kinetic energy of the particle, i.e.,  with high$-$energy molecular beams of constant velocity, the mass of the detected molecule. Moreover, silicon detectors are 100\% efficient and, owing to the kinematics,  small  detector sizes can cover 100\% of the solid angle for fragments emission. All the detectors are operating in coincidence, event by event. To get branching ratios on neutral species, the grid technique  \citep{PhysRevLett.26.602, Larsson1995403} may be used when the number of channels is small. In this method, a grid of known transmission is placed in front of the detector. Branching ratios of dissociation are linked to the recorded mass spectra through a set of linear equations. In the present work, because high$-$energy  molecular beams were used,  the shape of the current signal from  silicon detectors could  be analyzed to resolve a pile-up of several neutral fragments \citep{2002NIMPB.197..155C}. However, this technique was  insufficient to get the whole information for the  hydrogenated species. The grid technique was then mixed with a signal shape analysis to finally fully resolve the fragmentation of neutral and cationic C$_{\rm n}$, C$_{\rm n}$H, and C$_3$H$_2$ species \citep{2008JChPh.128l4312T}.

\subsection{High$-$velocity collision (HVC) processes}
\subsubsection{The excitation processes}

\begin{figure}
\begin{center}
\includegraphics[width=0.7\linewidth]{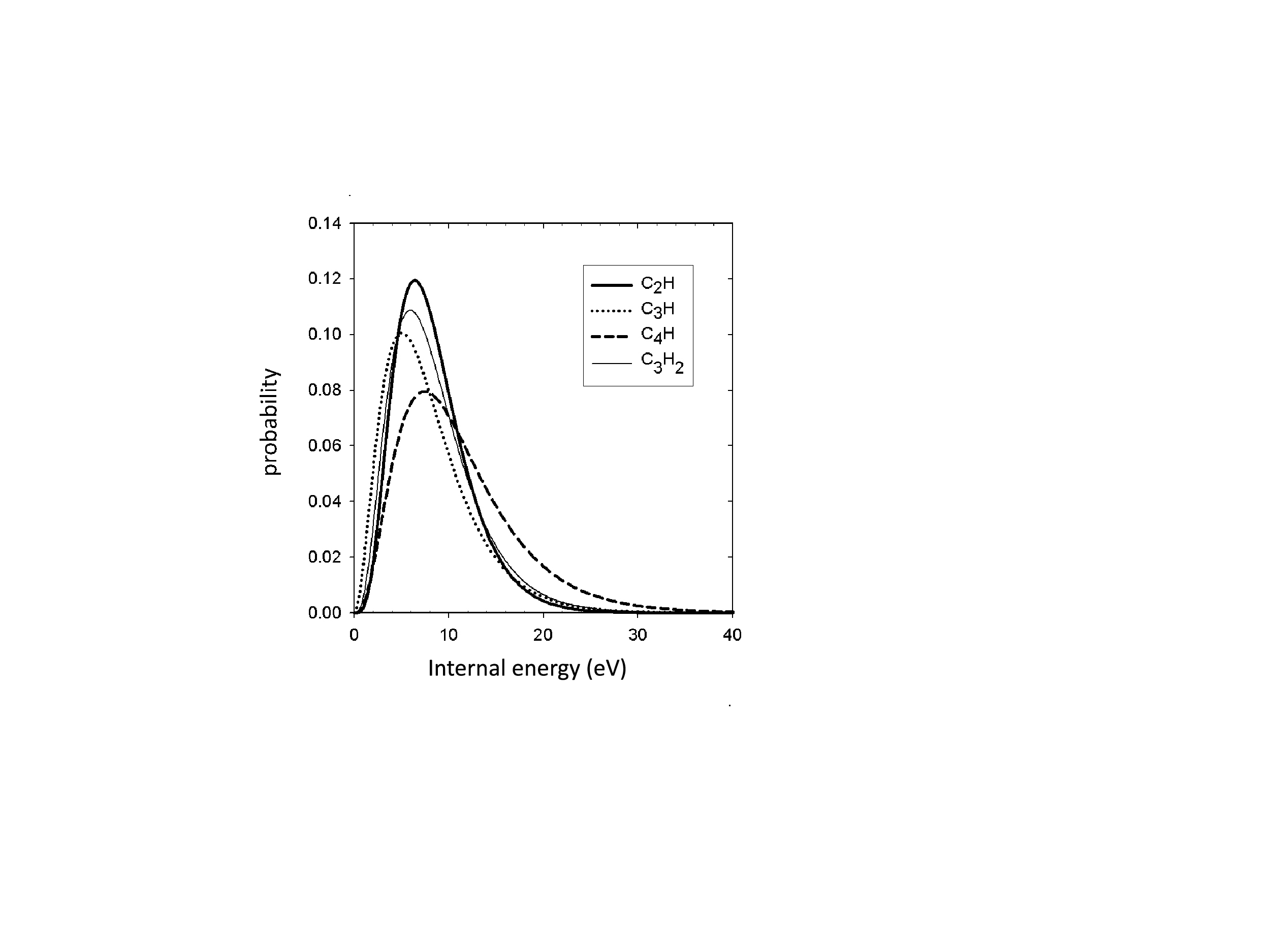}
\caption{Internal energy distributions following HVC-charge transfer (see text). Thick solid line: C$_2$H, dotted line: C$_3$H, broken line: C$_4$H, thin solid line: C$_3$H$_2$.  Peak energies and widths of distributions may be regarded with errors on the order of a few ev. Details on the method of extraction are given in \cite{2008JChPh.128l4312T}.}
\label{fig1}
\end{center}
\end{figure}

\begin{figure}
\begin{center}
\includegraphics[width=0.7\linewidth]{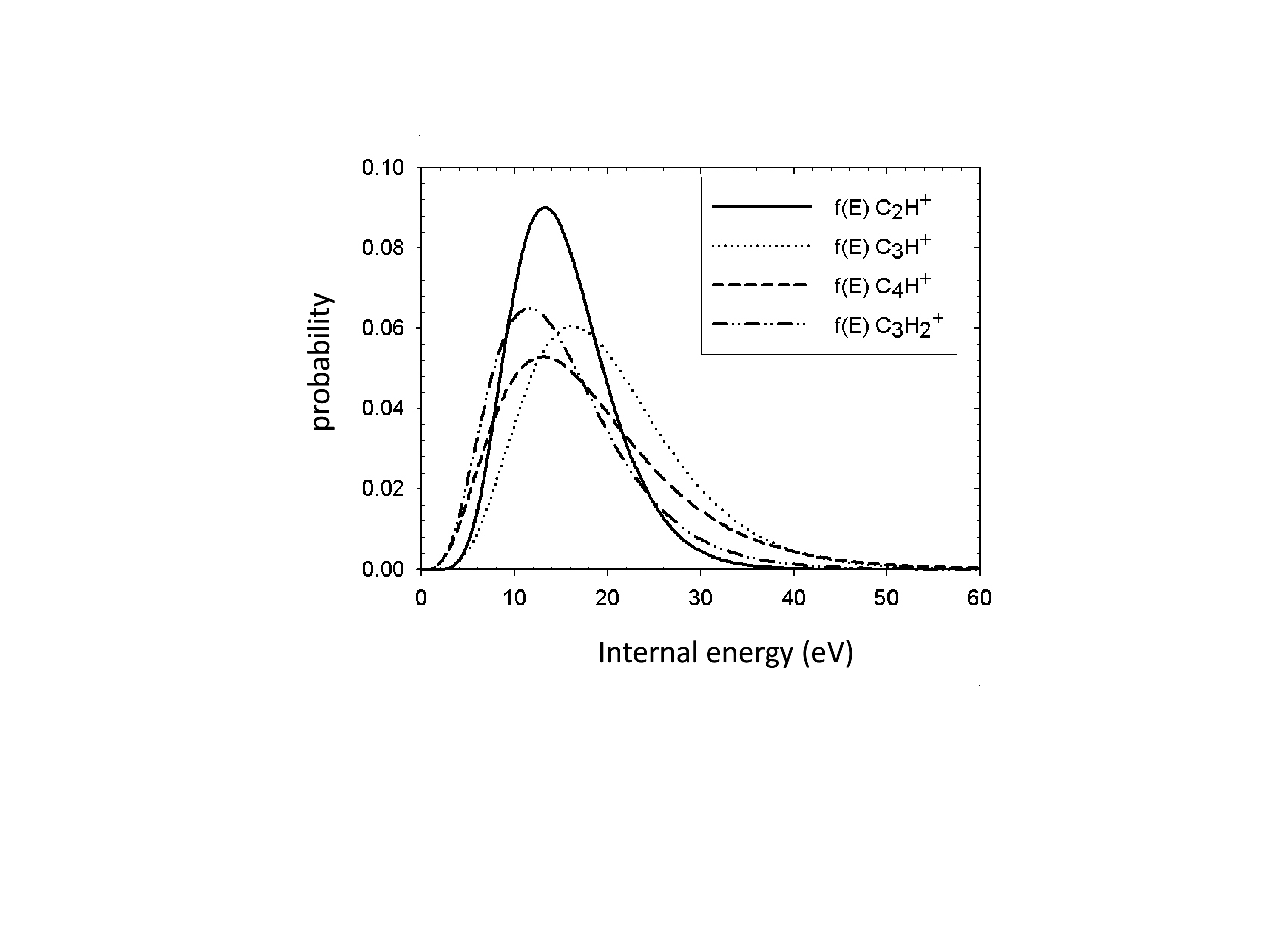}
\caption{Internal energy distributions following HVC-excitation (see text). Solid line: C$_2$H$^+$, dotted line: C$_3$H$^+$, broken line: C$_4$H$^+$, dot-dashed line: C$_3$H$_2^+$.  Peak energies and widths of distributions may be regarded with errors on the order of a few ev. Details on the method of extraction are given in \cite{2008JChPh.128l4312T}.}
\label{fig2}
\end{center}
\end{figure}

During the fast ($\sim 10^{-16}$ s) collision between a molecule X and an atom, charge transfer (HVC - CT) may occur:
\begin{equation}
\rm X^+ + He \rightarrow X^* + He^+
\end{equation}
Owing to the high initial velocity of the transferred electron, electronically excited states as well as the ground  state are populated \citep{2006JPhB...39.2593C}. Until now no calculations have been performed to predict the internal energy distribution associated with this process. Nevertheless, internal energy distributions may be deduced for the present species  from experimental multiplicity distribution, i.e. probabilities associated with a given number of emitted fragments \citep{2008JChPh.128l4312T}. Some internal energy distributions are reported in Fig.~\ref{fig1}.

In addition excitation (HVC - E) may also occur during the collision.
\begin{equation}
\rm X^+ + He \rightarrow X^{+*} + He
\end{equation}

Let us note in this velocity regime that only electronic excitations take place \citep{1998NIMPB.146...29W}. Calculations of internal energy following HVC-E have been performed through independent atom and electron model \citep{1993PhRvA..48.4784W} using doubly differential probabilities with the energy and impact parameter calculated within the classical trajectory monte carlo (CTMC) theory \citep{1999Maynard}. These calculated distributions agree very well with the energy distributions shown in Fig.~\ref{fig2},  wich are deduced from experimental multiplicity distributions.
	Ionization is also ocuring in HVC. In the present experiments with cationic beams it leads to multiply charged species which are generally also electronically excited \citep{PhysRevLett.104.043401}. We will not report on the fragmentation of these multiply charged species, because this is out of the scope of this paper.       

\subsubsection{Fragmentation of electronically excited species by high$-$velocity collision}

\begin{figure}
\begin{center}
\includegraphics[width=0.85\linewidth]{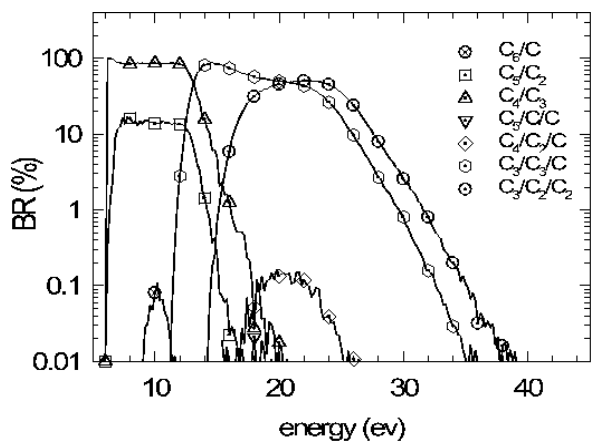}
\caption{ Theoretical fragmentation BRs as a function of internal excitation energy for a C$_7$ carbon cluster. Calculations were performed with the MMMC statistical theory \citep{2006IJMSp.252..126D}.}
\label{fig3}
\end{center}
\end{figure}

\begin{figure}
\begin{center}
\includegraphics[width=0.5\linewidth]{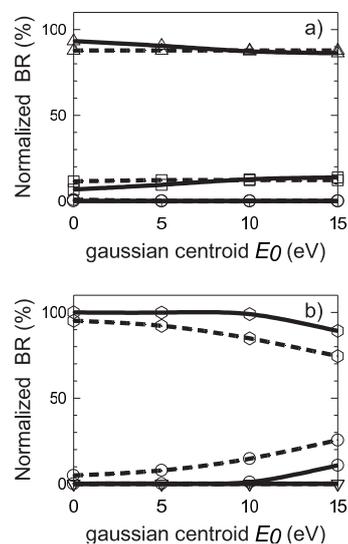}
\caption{C$_7$ calculated BRs  with various internal energy distribution. The BRs were obtained by convolution of theoretical curves of  Fig.~\ref{fig3} with the gaussian energy distribution: $\rm p(E) = \frac{1}{\sigma\sqrt{2\pi}}\exp(-\frac{1}{2}(\frac{E-E_0}{\sigma})^2)$. The BRs are normalized to the multiplicity probability. Solid lines: $\sigma$ = 2 eV dotted lines: $\sigma$ = 5 eV.a) BR for two fragments breakup: triangles up: C$_4$/C$_3$,   squares: C$_5$/C$_2$,   circle: C$_6$/C. b) BR for three fragments breakup:   hexagons: C$_3$/C$_3$/C,   circle: C$_3$/C$_2$/C$_2$,   triangle down: C$_4$/C$_2$/C and  C$_5$/C/C. }
\label{fig4}
\end{center}
\end{figure}

Statistical hypothesis has been proposed and often applied to calculate the fragmentation of finite size systems. It stipulates that all accessible micro-states are equiprobably populated, and conservations of energy and momentum define which these micro-states are. Their ensemble is forming a so-called phase space. Theoretical predictions have to rely on numeration of individual states or integration in the phase-space. Two physical situations are presumably close to the statistical behavior. In the first one, the system gets enough time to explore all the phase space whatever  the bottlenecks in this random exploration.  In the second physical situation,  the system has no time to explore the whole phase space, but  is prepared in such a large set of initial states that sampling of the phase space is presumably achieved by the entrance channels. Fragmentation of molecules that are electronically excited by HVC processes are clearly related to the second scenario and therefore can generally be interpreted in the frame of a statistical theory. For example, a statistical theory of fragmentation, the microcanonical metropolis monte carlo (MMMC) \citep{1995ZPhyD..35...27G}, has been applied to the carbon clusters C$_{\rm n}$ and found to agree very well with the experimental measurements \citep{2004PhRvL..93f3401M}.
 
For C$_{\rm n}$, C$_{\rm n}$H, and C$_3$H$_2$ the creation of each fragment costs roughly  the same amount of energy (4-7 eV \footnote{C$_3$/H$_2$ formation energy is only 3 eV, but there is a barrier close to 6 eV \citep{2008P&SS...56.1658L}.}). Therefore  the fragmentation BRs exhibit specific energy domains associated to a given multiplicity (see Fig.~\ref{fig3}). Normalizing the branching ratios to multiplicity probabilities somehow vanishes the role of the input energy on BRs. To illustrate this point, we present in Fig.~\ref{fig4} the calculated BRs for C$_7$. They were obtained by convolution of the theoretical curves of the Fig.~\ref{fig3} with Gaussian shapes centered at energies from 0 to 15 eV.The variations are found to be small, i.e. within $\pm$ 10$\%$ . It shows that as long as  phase space scrambling is achieved, normalized BRs are mostly insentive to the internal energy distribution, unlike the  multiplicity distributions.  

\subsection{Experimental results and analysis} 

\begin{figure}
\begin{center}
\includegraphics[width=1\linewidth]{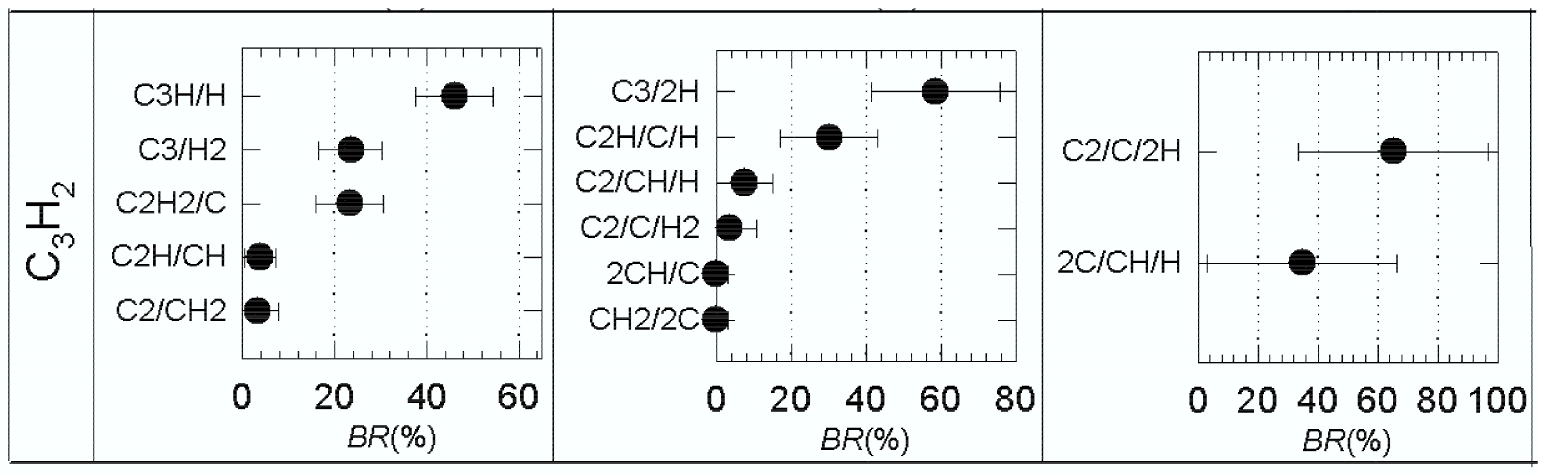}
\caption{Experimental branching ratios of neutral C$_3$H$_2$ produced in HVC-CT reaction at a fixed number of emitted fragments (N$_{\rm f}$) from left to right (N$_{\rm f}$) = 2, 3, 4.}
\label{fig5}
\end{center}
\end{figure}

\begin{figure}
\begin{center}
\includegraphics[width=1\linewidth]{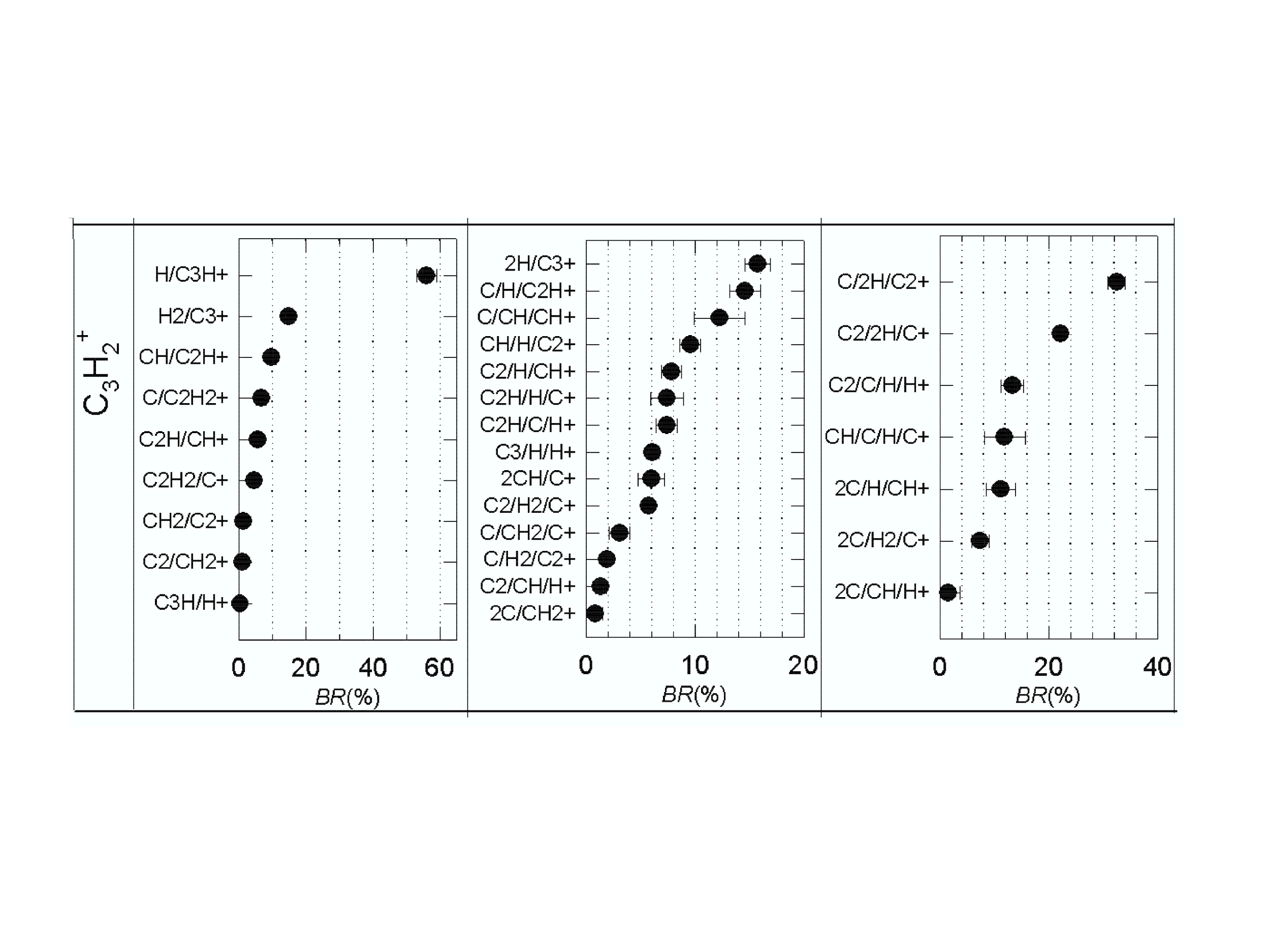}
\caption{Experimental branching ratios of C$_3$H$_2^+$ produced in HVC-E reaction at a fixed number of emitted fragments (N$_{\rm f}$) from left to righ (N$_{\rm f}$) = 2, 3, 4.}
\label{fig6}
\end{center}
\end{figure}

\begin{table}
\caption{Probability of  dissociation of C$_3$H$_2$ produced in HVC-CT reaction as a function of number of the emitted fragments (N$_{\rm f}$).N$_{\rm f}$=1 corresponds to non$-$dissociative HVC-CT.}
\begin{center}
\begin{tabular}{l|c}
\hline
\hline
N$_{\rm f}$ & Proba ($\pm$ err) \\
\hline
1 & 0.16 (0.05) \\
2 & 0.43 (0.07) \\ 
3 & 0.28 (0.08) \\
4 & 0.08 (0.03) \\
5 & 0.05 (0.01) \\
\hline
\end{tabular}
\end{center}
\label{Tab1}
\end{table}%

\begin{table}
\caption{Probability of  dissociation of (C$_3$H$_2^+$)* produced in HVC-E reaction as a function of number of the emitted fragments (N$_{\rm f}$). Non$-$dissociative excited species are not detected in the experiment. }
\begin{center}
\begin{tabular}{l|c}
\hline
\hline
N$_{\rm f}$ & Proba ($\pm$ err) \\
\hline
2 & 0.34 (0.02) \\ 
3 & 0.43 (0.03) \\
4 & 0.19 (0.08) \\
5 & 0.05 (0.01) \\
\hline
\end{tabular}
\end{center}
\label{Tab2}
\end{table}%

\begin{figure}
\begin{center}
\includegraphics[width=0.9\linewidth]{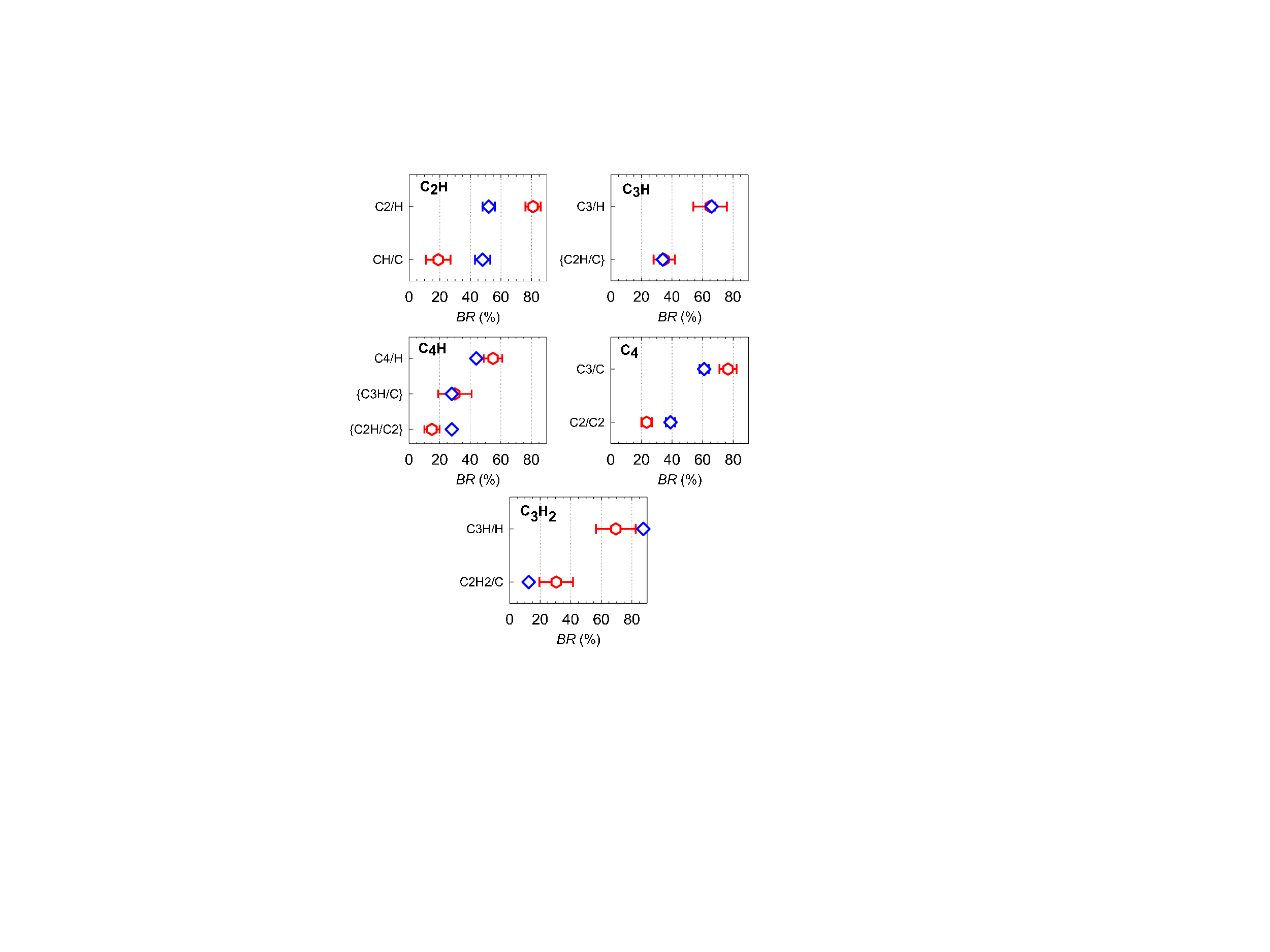}
\caption{Comparison between a two fragments breakup after dissociative recombination (DR) (blue diamonds)  and HVC-CT (red hexagons - this work). DR$-$BR are adapted from  \citet{2004PCCP....6..949E} for C$_2$H; \citet{Angelova20047} for C$_3$H, C$_3$H$_2$; \citet{Angelova2004195} for C$_4$H and  \citet{heber:022712} for C$_4$. The parenthesis on Y$-$axis labels mean that hydrogen may be localized on either of the two fragments in DR experiments.}
\label{fig7}
\end{center}
\end{figure}

\begin{figure}
\begin{center}
\includegraphics[width=0.9\linewidth]{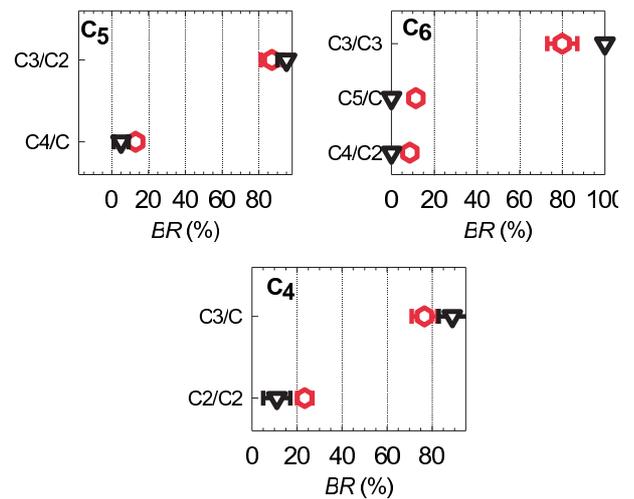}
\caption{Comparison between a two fragments$-$breakup after photodissociation (black triangles)  and HVC-CT (red hexagons - this work). The BRs for photodissociation are adapted from \cite{2000JPC} measurements. }
\label{fig8}
\end{center}
\end{figure}

\begin{figure}
\begin{center}
\includegraphics[width=0.9\linewidth]{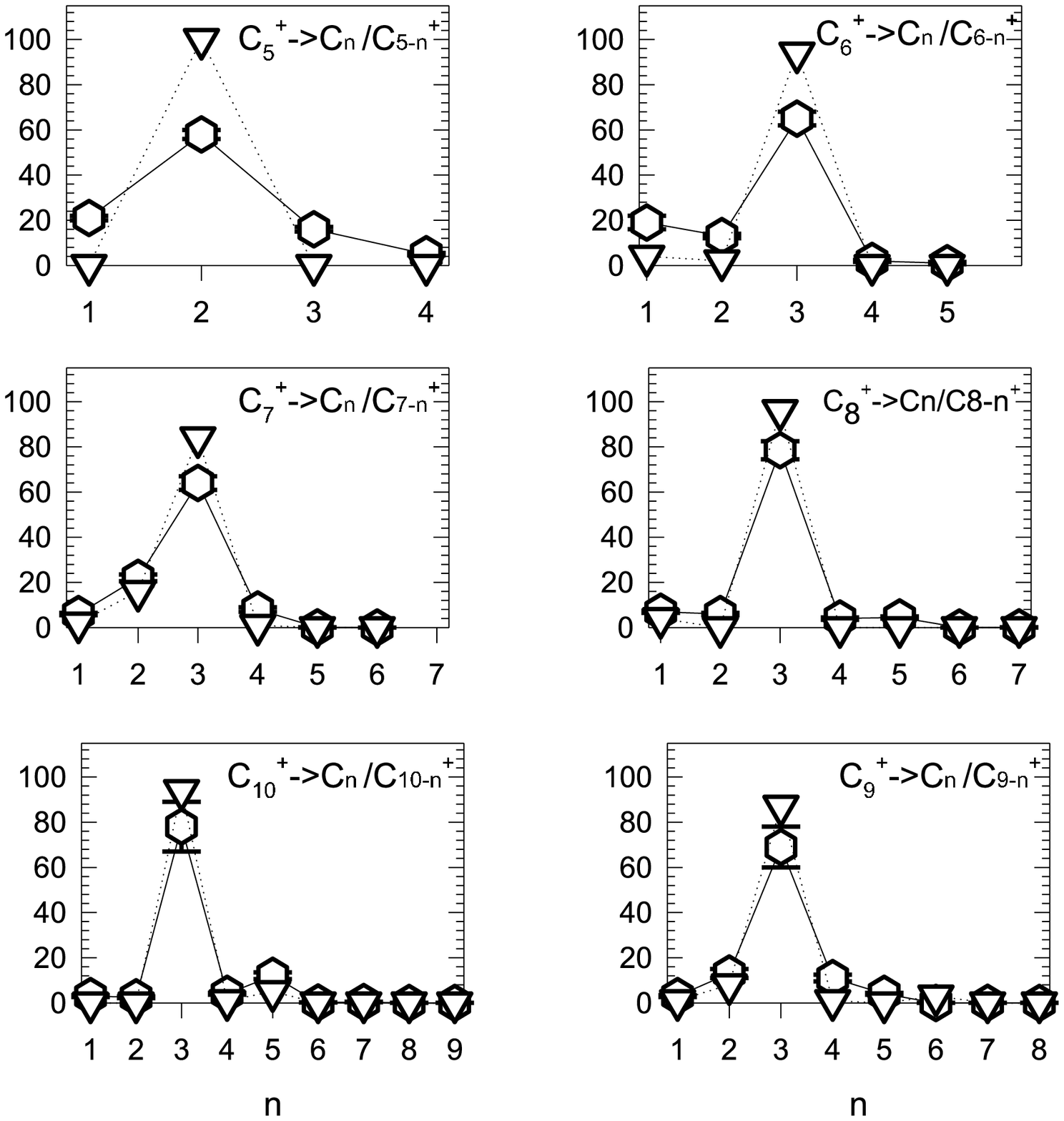}
\caption{Comparison between a two fragments$-$breakup after photodissociation (triangles)  and HVC-E (hexagons - this work). The BRs for photodissociation are adapted from \cite{1986ZPhyD...3..309G} }
\label{fig9}
\end{center}
\end{figure}

 Results for C$_3$H$_2$ fragmentation (multiplicity distribution: Table~\ref{Tab1}, BR: Fig.~\ref{fig5}) and C$_3$H$_2^+$ (multiplicity distribution: Table~\ref{Tab2}, BR: Fig.~\ref{fig6}) have been recently obtained. Detailed physical discussion on this species will be done elsewhere. Note that the detachment of one or two hydrogens is the most favorable channel. Fragmentation results for C$_{\rm n}$ and C$_{\rm n}$H species can be found elsewhere \citep{2004PhRvL..93f3401M,2006IJMSp.252..126D,2008JChPh.128l4312T}.A subset of these results are shown in Figs.~\ref{fig7} to   ~\ref{fig9}. Briefly, the most favorable channels were found to be H production for C$_{\rm n}$H and C$_3$ production for C$_{\rm n}$. These channels are indeed the most exothermic ones. For C$_{\rm n}$, C$_3$ is a magic number because of the shell closure \citep{VanOrden1998}. 

\section{Application of HVC fragmentation branching ratios in astrochemical reaction networks}

\subsection{Statistical fragmentation relevance in the context of ISM chemistry}

Three physical processes are involved in the creation of transient molecular electronically excited species in ISM \citep{1984inch.book.....D}. They are HVC with cosmic rays and secondary electrons \citep{1968ApJ...152..971S}, photoabsorption in the various local radiation fields \citep{1984inch.book.....D}, and recombination between molecular ions and thermal electrons in the local plasma \citep{2003guberman}. The two latter processes are generally dominating, except perhaps in dark clouds where CR$-$induced secondary electrons may have a significant contribution \citep{1983ApJ...267..603P,1987ApJ...323L.137G}.
 	   
Because of the energy and momentum conservation laws the internal energy, in the electronic recombination process,  should be close to the neutral species ionization potential (IP). In hydrocarbon molecules, the IP is always higher than the dissociation energy (DE). Then, fragmentation will occur rapidly for small size molecules, as compared to radiative deexcitation. According to the IP values \citep[8 to 12 eV;][]{VanOrden1998}   and DE \citep[4 to 7 eV;][]{2006IJMSp.252..126D,2008JChPh.128l4312T}, the channels that lead to two fragments are expected to be the most populated. 

Statistical fragmentation may be invoked in the dissociative recombination process because there are always many electronically excited states close to the IP in carbon and hydrocarbon molecules \citep{2008CP....343..292V}. As a result a rather efficient sampling of the phase space should be performed in the entrance channels, and dissociative recombination  BRs should be governed by a statistical fragmentation as in HVC. Figure 7 shows a comparison between HVC-CT and DR results for C$_4$, C$_2$H, C$_3$H, C$_4$H, and C$_3$H$_2$. Some of those results (C$_3$H and C$_4$) are not BR but a summation of BRs. Indeed, a resolution of C-H bond breaking is not always achieved in DR experiments nearby storage ring facilities due to the limitation of the grid method (see experimental section). The DR data agree with HVC-CT data within $\pm$ 10$\%$ on average. This small discrepancy arises because the internal energy distributions are quite different between HVC-CT (large distributions) and DR (narrow distribution around IP). An exception is seen in C$_2$H, where the agreement is poor. It may come from a non$-$statistical behavior of the fragmentation for a triatomic molecule this small.\\

In the ISM, molecules are photo-dissociated and photo-ionized because of the absorption of UV photons. These UV photons are produced by nearby stars. The interstellar standard radiation field (ISRF) embedding the gas in the Galaxy has been determined by Draine \citep{1978ApJS...36..595D}. In star$-$forming regions, the UV flux is the sum of the ISRF plus the emission spectra of the nearby stars. UV photons are absorbed in the continuum by dust and via discrete lines of the most important molecules. Photo-absorption rates thus depend on the optical depth. In the interior of dense clouds, UV photons are produced by the excitation in Lyman and Werner electronic states of molecular hydrogen by CR$-$secondary electrons \citep{1983ApJ...267..603P}.

Taking into account the strong density of electronically excited states for  the carbon and hydrocarbon molecules in the 6-11 eV energy range \citep{2008CP....343..292V}, many electronic transitions are expected to occur, resulting in a wide distribution of prepared excited states. Then fragmentation, as for HVC and DR, is expected  to be governed by statistical behavior. Figure 8 presents fragmentation BR for C$_4$, C$_5$, and C$_6$ carbon clusters obtained by photodissociation and those from HVC-CT. In the reported photodissociation experimental results \citep{2000JPC}, the incident photon energy has been varied up to 6 eV with some contribution of multi-photon absorption. The horizontal bars on these measurements correspond to a variation of the BRs with the different photon energies. One observes agreement within $\pm$ 10$\%$ between photodissociation and HVC-CT BR. Note that these photodissociation experiments were favorably compared to statistical calculations \citep{2000JPC}. Figure 9 presents the same comparison for cationic C$_n^+$ (n= 5 to 10) species. Photodissociation was produced by single and runaway multiphoton absorption \citep{1991JChPh..95.4719S,1986ZPhyD...3..309G}. Therefore, those results have to be considered with caution. Inside the above restrictions, the BRs from photodissociation and HVC-E processes still agree also quite well.
   
	In view of the above comparisons, a universality of statistical BRs  may be proposed. In order to take into account the particularity of the different electronic processes,  an error bar (1 $\sigma$) of $\pm$ 10$\%$ seems reasonable to apply.

\subsection{New BRs for astrochemistry databases}
 
 \begin{table}
\caption{New HVC-SUBR for CR reactions. Total rates and BR of OSU-01-2007 ( http:www.eric/o.html) are also reported. We do not report on C$_{10}$ because this reaction is not included in OSU-01-2007.}
\begin{center}
\begin{tabular}{c|p{1.5cm}|p{1.4cm}|p{1.0cm}|p{1.2cm}}
\hline
\hline
Reactants & Total rate (10$^{3}$ s$^{-1}$) (OSU-01-2007) & Products & HVC-SUBR & OSU-01-2007-BR \\
\hline
C$_2$H + cr & 5.0 & C$_2$ + H &  0.81 & 1.00  \\
& & C+CH & 0.19 & 0.00  \\
\hline
C$_3$H + cr & 5.0 & C$_3$ + H & 0.65&  1.00  \\
& & C$_2$H + C &  0.33 & 0.00  \\
&  & C$_2$ + CH & 0.02&  0.00  \\
\hline
C$_4$ + cr &  1.0 & C$_3$ + C & 0.77 & 1.00  \\
& & C$_2$ + C$_2$ & 0.23 & 0.00  \\
\hline
C$_3$H$_2$ + cr & 5.0  & C$_3$H + H & 0.46 & 1.00  \\
& & C$_3$ + H$_2$ & 0.24 & 0.00  \\
& & C$_2$H$_2$ + C & 0.23 & 0.00  \\
& & C$_2$H + CH &  0.04 & 0.00  \\
& & C$_2$ + CH$_2$ & 0.03 & 0.00  \\
\hline
C$_4$H + cr & 5.00 & C$_4$ + H & 0.58 & 1.00  \\
& & C$_3$H + C &  0.26&  0.00  \\
 & & C$_3$ + CH & 0.00 & 0.00  \\
& & C$_2$H + C$_2$ & 0.16& 0.00  \\
\hline
C$_5$ + cr & 1.00 & C$_4$ + C & 0.13 & 1.00  \\
& & C$_3$ + C$_2$ & 0.87 & 0.00  \\
\hline
C$_6$ + cr & 1.00 & C$_5$ + C & 0.09 & 1.00  \\
& & C$_4$ + C$_2$ & 0.11 & 0.00  \\
& & C$_3$ + C$_3$ & 0.80 & 0.00  \\
\hline
C$_7$ + cr & 1.00 & C$_6$ + C & 0.01 & 1.00  \\
& & C$_5$ + C$_2$ &  0.19 & 0.00  \\
& & C$_4$ + C$_3$ & 0.80&  0.00  \\
\hline
C$_8$ + cr & 1.00  & C$_7$ + C & 0.03  &1.00  \\
& & C$_6$ + C$_2$ & 0.01 & 0.00  \\
& & C$_5$ + C$_3$ & 0.90 & 0.00  \\
& & C$_4$ + C$_4$ & 0.06 & 0.00  \\
\hline
C$_9$ + cr & 1.00 & C$_8$ + C & 0.00 & 1.00  \\
& & C$_7$ + C$_2$ & 0.06 & 0.00  \\ 
& & C$_6$ + C$_3$ & 0.66&  0.00  \\
& & C$_5$+C$_4$ & 0.28 & 0.00  \\
\hline
\end{tabular}
\end{center}
\label{tab3}
\end{table}

 \begin{table}
\caption{New HVC-SUBR for DR reactions. Total rates and BR of OSU-01-2007 ( http:www.eric/o.html) are also reported. We do not report about DR on C$_2$H$^+$ and C$_4^+$  because OSU-01-2007 used experimental measurements.}
\begin{center}
\begin{tabular}{c|p{1.5cm}|p{1.4cm}|p{1.0cm}|p{1.2cm}}
\hline
\hline
Reactants & Total rate (10$^{-7}$ cm$^{3}$ s$^{-1}$) (OSU-01-2007) & Products & HVC-SUBR & OSU-01-2007-BR \\
\hline
C$_3$H$^+$ + e$^-$&	3.0	&C$_3$+H	&0.65&	0.50\\
	&&	C$_2$H+C	&0.33&	0.50\\
		&& C$_2$ + CH	&0.02&	0.00\\
\hline
C$_3$H$_2^+$ + e$^-$ &	3.6	&C$_3$H + H&	0.46&	0.42\\
	&&	C$_3$ + H$_2$&	0.23&	0.42\\
	&&	C$_2$H$_2$ + C&	0.24&	0.08\\
	&&	C$_2$H + CH	&0.04&	0.00\\
	&&	C$_2$ + CH$_2$&	0.04	&0.08\\
\hline
C$_4$H$^+$ + e$^-$&	3.00&	C$_4$ + H&	0.58	&0.40\\
	&&	C$_3$H + C	&0.26&	0.15\\
	&&	C$_3$ + CH	&0.00&	0.15\\
	&&	C$_2$H + C$_2$&	0.16&	0.30\\
\hline
C$_5^+$ + e$^-$&	3.00&	C$_4$ + C	&0.13&	0.50\\
	&&	C$_3$ + C$_2$&	0.87	&0.50\\
\hline
C$_6^+$ + e$^-$&	20.00&	C$_5$ + C&	0.09&	0.50\\
	&&	C$_4$ + C$_2$&	0.11	&0.50\\
	&&	C$_3$ + C$_3$&	0.80&	0.00\\
\hline
C$_7^+$ + e$^-$&	23.00	&C$_6$ + C	&0.01&	0.43\\
	&&	C$_5$ + C$_2$	&0.19&	0.43\\
	&&	C$_4$ + C$_3$&	0.80	&0.14\\
\hline
C$_8^+$ + e$^-$	&20.00&	C$_7$ + C&	0.03	&0.50\\
	&&	C$_6$ + C$_2$ &	0.01&	0.50\\
	&&	C$_5$ + C$_3$	&0.90&	0.00\\
	&&	C$_4$ + C$_4$	&0.06	&0.00\\
\hline
C$_9^+$ + e$^-$ &	20.00	&C$_8$ + C	&0.00&	0.50\\
	&&	C$_7$ + C$_2$ &	0.06	&0.50\\
	&&	C$_6$ + C$_3$&	0.66	&0.00\\
	&&	C$_4$ + C$_4$&	0.28&	0.00\\
\hline
C$_{10}^+$ + e$^-$&	20.00	&C$_9$ + C&	0.01	&0.50\\
	&&	C$_8$ + C$_2$	&0.01	&0.50\\
	&&	C$_7$ + C$_3$&	0.70	&0.00\\
	&&	C$_6$ + C$_4$&	0.03	&0.00\\
	&&	C$_5$ + C$_5$ &	0.25&	0.00\\
\hline
\end{tabular}
\end{center}
\label{tab4}
\end{table}%

 \begin{table}
\caption{New HVC-SUBR for photo dissociation reactions. Total rates and BR of OSU-01-2007 (http:www.eric/o.html) are also reported.}
\begin{center}
\begin{tabular}{c|p{1.5cm}|p{1.4cm}|p{1.0cm}|p{1.2cm}}
\hline
\hline
Reactants & Total rate (10$^{-9}$ s$^{-1}$) (OSU-01-2007) & Products & HVC-SUBR & OSU-01-2007-BR \\
\hline
C$_2$H + h$\nu$	& 1.0	&C$_2$ + H&	0.81	&1.00 \\
	&&	C + CH&	0.19&	0.00 \\
\hline
C$_3$H + h$\nu$	&1.0	& C$_3$  + H &	0.65&	1.00 \\
	&&	C$_2$H + C	&0.33&	0.00 \\
	&&	C$_2$ + CH	&0.02	&0.00 \\
\hline
C$_4$ + h$\nu$&	0.4	&C$_3$ + C	&0.77&	0.50 \\
	&&	C$_2$ + C$_2$ &	0.23	&0.50 \\
\hline
C$_3$H$_2$ + h$\nu$	&2.9	& C$_3$H + H&	0.46	&0.66 \\
	&&	C$_3$ + H$_2$	&0.23&	0.34 \\
	&&	C$_2$H$_2$ + C&	0.24&	0.00 \\
	&&	C$_2$H + CH &	0.04&	0.00 \\
	&&	C$_2$ + CH$_2$	&0.04&	0.00 \\
\hline
C$_4$H + h$\nu$	&2.0	&C$_4$ + H&	0.58	&0.50 \\
	&&	C$_3$H + C &	0.26&	0.00 \\
	&&	C$_3$ + CH&	0.00&	0.00 \\
	&&	C$_2$H + C$_2$ & 	0.16	&0.50 \\
\hline
C$_5$ + h$\nu$ &	0.01	&C$_4$ +  C&	0.13	&0.00 \\
		&&C$_3$ + C$_2$&	0.87	&1.00 \\
\hline
C$_6$ + h$\nu$&	1.0	&C$_5$ + C	&0.09	&1.00 \\
	&&	C$_4$ + C$_2$	&0.11	&0.00 \\
	&&	C$_3$ + C$_3$	&0.80	&0.00 \\
\hline
C$_7$ + h$\nu$	&1.0	& C$_6$ + C	&0.01&	1.00 \\
	&&	C$_5$ + C$_2$&	0.19	&0.00 \\
	&&	C$_4$ + C$_3$	&0.80	&0.00 \\
\hline
C$_8$ + h$\nu$	& 1.0	&C$_7$ + C&	0.03	&1.00 \\
	&&	C$_6$ + C$_2$ &	0.01	&0.00 \\
	&&	C$_5$ + C$_3$ &	0.90&	0.00 \\
	&&	C$_4$ + C$_4$ &	0.06&	0.00 \\
\hline
C$_9$ + h$\nu$&	1.0	& C$_8$ + C	&0.00&	1.00 \\
	&&	C$_7$ + C$_2$	&0.06	&0.00 \\
	&&	C$_6$ + C$_3$	&0.66	&0.00 \\
	&&	C$_5$ + C$_4$ &	0.28	&0.00 \\
\hline
C$_{10}$ + h$\nu$	&1.14	& C$_9$ + C	&0.01&	0.82 \\
	&&	C$_8$ + C$_2$ &	0.01	&0.17 \\
	&&	C$_7$ + C$_3$ &	0.70	&0.00 \\
	&&	C$_6$ + C$_4$ &	0.03&	0.01 \\
	&&	C$_5$ + C$_5$ &	0.25&	0.00 \\
\hline
\end{tabular}
\end{center}
\label{tab5}
\end{table}%

We propose here, to use the present complete set of BR obtained with HVC as statistical universal BRs (HVC-SUBR) for C$_n$, C$_n$H, and C$_3$H$_2$ when they are missing or guessed in ISM databases. Table~\ref{tab3} presents CR HVC-SUBR for C$_n$, C$_n$H, and C$_3$H$_2$ together with OSU-01-2007 CR-BR. We do not report the Umist data base values because the BR are almost the same as those of OSU-01-2007. When reactions were not considered in OSU-01-2007, we do however used Umist06 rates.  Note that the most recent version of the osu database (USU.2009) has the same BR as OSU-01-2007 for these species.  All CR-BR from OSU-01-2007 result from the pioneering estimates \citep{1984ApJS...56..231L}. In most cases, they used  zero level statistical behavior  to predict BR,  i.e.  all the dissociation goes only to the most exothermic channel.  It is consistent with the  HVC-SUBR obtained for the C$_n$H,  but not for the C$_n$ species. Indeed, they assumed C$_{n-1}$/C and C$_{n-1}$/C$_2$ to be the most exothermic channels, whereas it is always the C$_{n-3}$/C$_3$ channel. In general, HVC-SUBR lead to a much wider dispersion of the fragmentation channels. It is remarkable that C$_3$ production by CR is strongly enhanced. 
 
 Table~\ref{tab4} presents new DR HVC-SUBR for C$_n^+$, C$_n$H$^+$, and C$_3$H$_2^+$  together with OSU-01-2007 BR and reaction rates. We do not report on C$_2$H$^+$ and C$_4^+$ because OSU-01-2007 used experimental DR-BR \citep{2004PCCP....6..949E, heber:022712}. For C$_3$H$^+$, in spite of existing partial measurements \citep{Angelova20047}, the  OSU-01-2007 database uses the same probability for the emission of C or H. Although \cite{1984ApJS...56..231L} assumed    100\% of H emission, equiprobability  was proposed  by  \cite{1988A&A...194..250M}, based on more detailed calculations. The HVC-SUBR are indeed between these two extreme situations. For C$_4$H$^+$, OSU-01-2007 uses the experimental DR- BR \citep{Angelova2004195} for the breaking of the C-C bonds and, without any available information on the ratio between emission of C or CH, OSU-01-2007 took it equal to 1. The HVC-SUBR results show that actually CH emission is very unlikely. For C$_3$H$_2^+$, OSU-01-2007 uses the experimental DR-BR \citep{Angelova20047} for the C-C bonds. For the missing experimental information on the proportion between H and H$_2$ emission when the carbon skeleton stays intact and the proportion between C and CH when it is broken, OSU-01-2007 still assumed equal probability. As for C$_3$H$^+$, HVC-SUBR show that CH emission is unlikely. For the proportion of evaporation of H and H$_2$, equal probability is not too far from HVC-SUBR. It is remarkable  that the HVC-SUBR give a ratio H$_2$/H  (35\%) in qualitative agreement  with the recent experiment of neutral-neutral  reaction C+C$_2$H$_2$ $\rightarrow$ C$_3$H$_2^*$ \citep{2008P&SS...56.1658L}.  For the C$_n^+$ species, as for CR, note that HVC-SUBR predict an enhanced production of C$_3$ driven by DR together with a more extended dispersion in the carbon cluster$-$mass daughters  compared to OSU-01-2007.  
 
	Table~\ref{tab5} presents a new photon HVC-SUBR for C$_n$, C$_n$H, and C$_3$H$_2$ together with OSU-01-2007 BR and reaction rates. Note that the OSU database has been developed mainly for cold dark clouds. Therefore, references about  BR for photodissociation processes are scarce. OSU-01-2007 Photodissociation BR for C$_2$H and C$_3$H  used zero level statistical picture: only H emission is allowed.  For C$_4$H, it assumed equal probability between emission of C$_2$ and H. For all C$_n$H species HVC-SUBR predict H emission to be by far the dominant channel. For C$_3$H$_2$, OSU-01-2007 allows H or H$_2$ emission.  Emission of C should also be an open channel. For the C$_n$ series, as for the two previous processes, OSU-01-2007 assumed exclusive C emission. Exception arises from C$_{10}$ where RRKM statistical calculations were performed \citep{1995IJMSI.149..321B}.  Again HVC-SUBR predicts a stronger enhancement of the C$_3$ cluster production compared to OSU-01-2007.

\section{Effects of the new branching ratios on chemical model predictions for dense clouds}

\subsection{Dense cloud model} 

In order to test the effect of the new branching ratios on chemical model predictions for dense clouds, we used the Nahoon chemical model developed by V. Wakelam \citep[][http://www.obs.u-bordeaux1.fr/amor/VWakelam]{2004A&A...422..159W}. This model follows the time-dependent chemistry of gas-phase species at a fixed temperature and density. We  used Nahoon  for a single spatial point (0D). We considered "typical" dense cloud conditions: a temperature of 10~K, an H density of $2\times 10^4$~cm$^{-3}$, a visual extinction (Av) of 10 and a cosmic-ray ionization rate of $1.3\times 10^{-17}$~s$^{-1}$. For the initial conditions, we started with all the elements but H in the atomic form with the low-metal elemental abundances from \citet{1982ApJS...48..321G}. The standard network used for this analysis is OSU-01-2007 (http://www.physics.ohio-state.edu/$\sim$eric/research.html), which contains 452 species and 4430 reactions. We updated the branching ratios listed in Tables~\ref{tab3}, \ref{tab4}, and \ref{tab5} for 1) cosmic-ray dissociations of C$_n$ (n = 4 to 9), C$_n$H (n = 2 to 4) and C$_3$H$_2$, 2) dissociative recombinations of C$_n^+$ (n = 5 to 10), C$_n$H$^+$ (n = 3 to 4) and C$_3$H$_2^+$,  and 3) photo-dissociations of C$_n$ (n = 4 to 10), C$_n$H (n = 2 to 4) and C$_3$H$_2$. We then let the system evolve over $10^8$~yr, when it reaches steady-state.

\subsection{Results}  

\begin{figure}
\begin{center}
\includegraphics[width=0.9\linewidth]{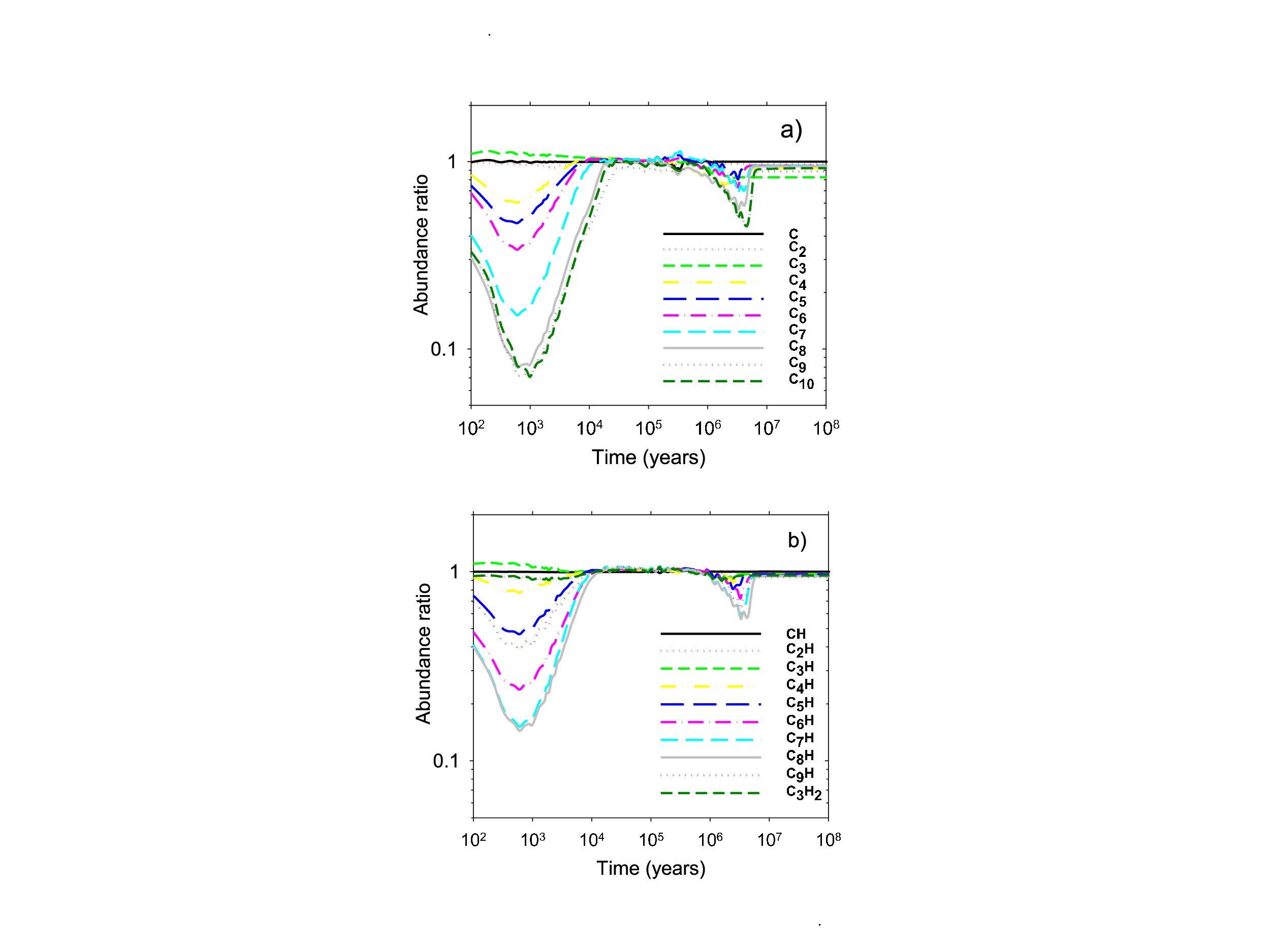}
\caption{Evolution with time of  the ratios of abundances computed with the updated database and with the standard network (OSU-01-2007) for a dense clouds for a) C$_n$  and b)  C$_n$H$_m$ species. (see text) }
\label{fig10}
\end{center}
\end{figure}

Figure~\ref{fig10} shows the ratio between carbon$-$bearing species abundances computed with the updated database and those computed with the standard network OSU-01-2007. The new branching ratios modify the species abundances at two different times. The first one is a very early stage, before $10^4$~yr, which is irrelevant for dense cloud chemistry. The second is much later, between $10^6$ and $10^7$~yr, but then the effect of the new branching ratios is less important. At the typical dense cloud age of $10^5$~yr \citep{2006A&A...451..551W}, the new branching ratios are unimportant. Higher ages have however been found for TMC-1 with other models \citep{1998ApJ...501..207T}, so the period around $10^6$~yr is not without interest. After $10^6$~yr, the maximum effect is obtained for the largest molecules. The C$_n$H molecules are an exception because C$_7$H and C$_8$H are more influenced than C$_9$H. All the C-bearing species abundances are decreased by the new BR by a maximum factor of two. The weak sensitivity of TMC-1 chemistry to BR has already been pointed out \citep{1988A&A...194..250M,1997ApJ...485..689H}   \\

\section{Effects of the new branching ratios on chemical model predictions for photon$-$dominated regions}  


\subsection{PDR model} 

We tested the influence of these new rates on a PDR model. For this, we used the Meudon PDR code (http//pdr.obspm.fr) described in \citet{2006ApJS..164..506L}. The Meudon PDR code computes the structure of a 1D plan-parallel and stationary slab of dust and gas. It consistently solves  the radiative transfer from far UV to sub-millimeter, chemistry, and thermal balance. To test the influence of the new rates, we reproduced the model of the Horse Head by this PDR code presented in  \citet{2005A&A...435..885P} first with an old chemical network based on the rates provided by OSU-01-2007, secondly with the new branching ratios presented in this paper. The Horse Head is good candidate because of the large number of observed hydrocarbons. Pety et al. (2005) suggest that fragmentation of PAHs can contribute to the synthesis of small hydrocarbons and conclude by mentioning the need for precise chemical rates to perform chemical models in these regions. The proton density in the Horse Head is estimated to n$_{\textrm{H}}$ = 10$^5$ cm$^{-3}$ and the intensity of the incident UV flux to 100 times the ISRF \citep[in Draine's units,][]{1978ApJS...36..595D}. We adopted a flux of comic rays of $5\times10^{-17}$ per H and per second. The model assumes a semi-infinite cloud. Our objective is to compare the profile of abundances of some relevant species computed with our new branching ratios and with older ones. To refine models of the Horse Head is beyond the scope of this paper.

\subsection{Results}   

\begin{figure}
\begin{center}
\includegraphics[width=0.9\linewidth]{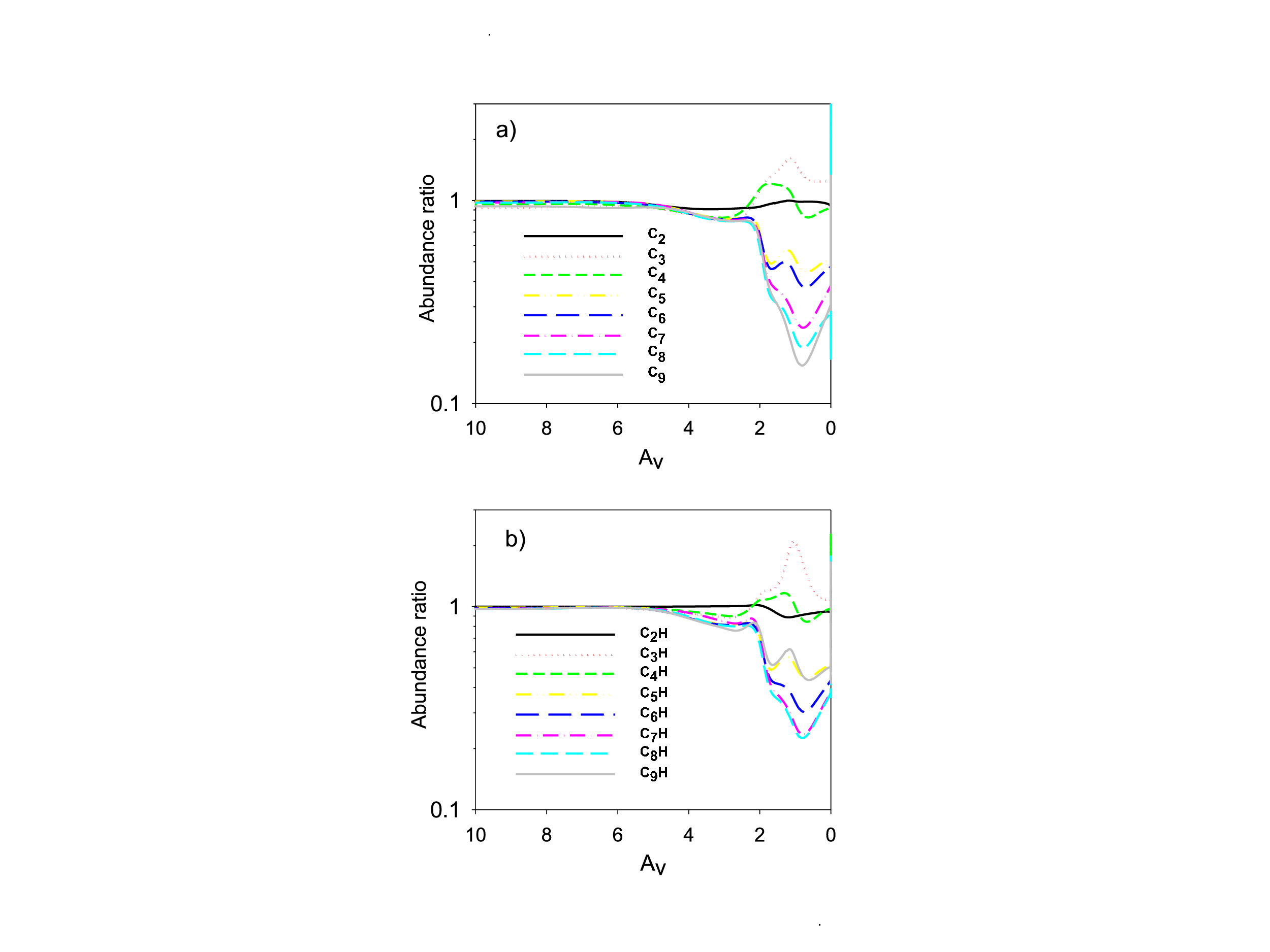}
\caption{Evolution, with the \textbf{visual} extinction Av, of the ratios of aboundances computed with the updated database and with the standard network (OSU-01-2007) for a PDR a) C$_n$  and b)  C$_n$H$_m$ species (see text). }
\label{fig11}
\end{center}
\end{figure}

Figure \ref{fig11} presents the ratios of abundances provided by the two chemistries of C$_{n}$ and C$_{n}$H with $n$ from 2 to 9 as a function of the position in the cloud expressed in visual extinction. This ratio is defined as the abundance provided with the new chemistry divided by the abundance obtained with a older rates. First we note that the effect of the new rates are only visible for A$_{\textrm{V}}$ lower than 4. Indeed, it is in this region that the dissociative recombinations and photo-processes dominate the other chemical processes.\\

It is often difficult to reproduce the abundance of C$_3$ in PDR models (for example in the diffuse interstellar gas towards Zeta Persus, the model by \citet{2004A&A...417..993L} requires a high density component to reproduce this molecule). Figure \ref{fig11} shows that the new branching ratios enhance the abundance of this molecule. This is explained by two factors. First, the new branching ratios of the C$_3$H$^+$ recombination reaction enhance the route leading to C$_3$ by 30\%. Secondly, we showed that the dissociative recombination of C$_6^+$ can efficiently produce  two C$_3$ molecules.This route was not considered in previous chemistries. \\
For C$_n$ molecules with $n>$4, the new branching ratios systematically produce a decrease in the abundances. 
The model shows that the abundance of C$_3$H is significantly enhanced in the photodissociation area (Fig. \ref{fig11}b). This is explained by the new photodissociation route of C$_4$H. This route was not considered in the OSU database.\\

Finally, the abundances of the hydrocarbons, C$_n$H molecules (n$>$3), are reduced compared to the old chemistry. This is directly linked to the decrease in the abundances the C$_n$ molecules. Indeed, the chain of $  $reactions leading to the hydrocarbons with n$>$3 is 

\begin{equation}
\rm C_n\stackrel{C^+}{\rightarrow}C_{n+1}^{+}\stackrel{H_2}{\rightarrow}\textrm{C}_{n+1}H^+\stackrel{e^-}{\rightarrow}\textrm{C}_nH.
\end{equation}

Because the abundance of C$_n$ molecules is reduced by the new branching ratios, the abundances of the C$_n$H are also reduced.\\

\section{Conclusions}  
  
High velocity collision in inverse kinematics scheme was used to measure the complete fragmentation pattern of electronically excited C$_{n}$ ($n$=2 to 10), C$_{n}$H ($n$=2 to 4) and C$_3$H$_2$ molecules. Branching ratios of dissociation were deduced from those experiments. The comparison between the branching ratios obtained in this work and other types of experiments showed a good agreement. It was interpreted as the signature of a statistical behavior of the fragmentation. We thus propose new branching ratios for: 1) the dissociation of C$_n$ ($n$=2-10), C$_n$H ($n$=2-4) and C$_3$H$_2$ molecules by interstellar UV photons or cosmic-ray processes and 2) the dissociative recombination of C$_n^+$ ($n$=5-10), C$_n$H$^+$ ($n$=3-4) and C$_3$H$_2^+$. The new values have been tested in dense cloud and PDR models. We showed that only chemistry occurring at A$_{\rm V}$ smaller than 4 is really affected. We however recommend astrochemists to use these branching ratios, even for dense chemistry, because it is well known that the importance of a reaction depends on the network we use. The data published in this paper have been added to the online database KIDA (KInetic Database for Astrochemistry, http://kida.obs.u-bordeaux1.fr).

\begin{acknowledgements}
The national CNRS program PCMI (Physique et Chimie du Milieu Interstellaire), the CNRS IN2P3 (Institut National de Physique Nucl\'eaire et de Physique des Particules) and the PPF Mati\`ere Carbon\'ee of the University Paris Sud 11 are acknowledged for their financial support.

\end{acknowledgements}

\end{document}